# Second harmonic generation of spatiotemporal optical vortices (STOVs) and conservation of orbital angular momentum

S. W. Hancock, S. Zahedpour, H. M. Milchberg[†]

*Institute for Research in Electronics and Applied Physics, University of Maryland, College Park, Maryland, 20742, USA*
[†]*milch@umd.edu*

**Spatiotemporal optical vortices (STOVs) are a new type of intrinsic optical orbital angular momentum (OAM) structure in which the OAM vector is orthogonal to the propagation direction [Optica 6, 1547, (2019)] and the optical phase circulates in space-time. Here, we experimentally and theoretically demonstrate, for the first time, the generation of the second harmonic of a STOV-carrying pulse along with the conservation of STOV-based OAM. Our experiments verify that individual photons can have intrinsic orbital angular momentum perpendicular to their propagation direction.**

A spatiotemporal optical vortex (STOV) [1,2] is an electromagnetic structure with orbital angular momentum (OAM) and optical phase circulation defined in space-time and is supported by a polychromatic pulse [3]. For a STOV-carrying pulse propagating in free space [2], the OAM vector is perpendicular to the direction of propagation. This contrasts with a conventional space-defined optical vortex, which can be supported by a monochromatic beam, and where the OAM vector is parallel/anti-parallel to the direction of propagation and the optical phase winding is in the plane transverse to propagation [4-6]. Examples of the latter include Bessel-Gauss ($BG_l$) or Laguerre-Gaussian ($LG_{pl}$) modes with nonzero azimuthal index $l$ [4]. STOVs were first measured as naturally emergent from filamentation processes in material media [1] and can be constructed using a $4f$ pulse shaper, as originally proposed in [7], with free-space STOV propagation first demonstrated in [2, 8] and later confirmed by [9].

In second harmonic generation ($\omega \rightarrow 2\omega$) of conventional OAM beams, the second harmonic photons carry twice the OAM of fundamental beam photons ($l\hbar \rightarrow 2l\hbar$) [10-14], where $l$ is the OAM quantum number or beam topological charge. Similarly, in sum or difference frequency generation, the OAM of two fundamental modes add [15]. In the case of $q^{th}$ order high harmonic generation with a mode of charge $l$, the resulting photons have OAM $ql\hbar$ [16-19]. The conservation of conventional OAM under these wide conditions has prompted our measurements of harmonic generation and OAM conservation in nonlinear interactions of STOV-carrying pulses, as first presented in [20,21].

In this paper we demonstrate, for the first time, the second harmonic generation (SHG) of STOVs and conservation of STOV orbital angular momentum. Because SHG is fundamentally an interaction process of the quantized electromagnetic field, and because all photons in the STOV pulse from our pulse shaper carry the same bandwidth, polarization, and spatiotemporal phase, our results verify that individual photons can have OAM orthogonal to their direction of propagation. To perform the measurements, we use a single shot measurement technique [2,22] that captures the fundamental and SHG amplitude and phase structure in mid-flight. Accompanying the measurements are simulations exploring the conversion process and the propagation of STOVs in material media.

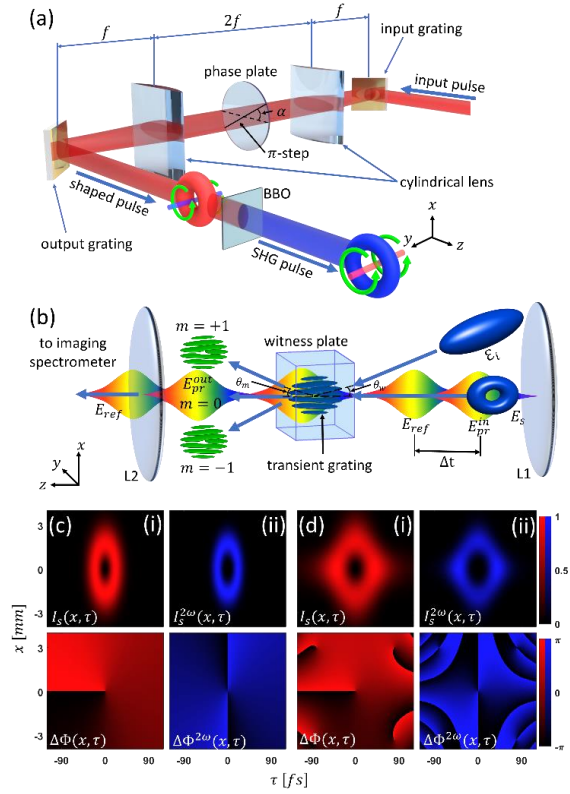

**Fig. 1. (a)** $4f$ pulse shaper used to generate a $l = \pm 1$ STOV –carrying pulse at 800nm, composed of two 1200 groove/mm gratings, two $f = 10$ cm cylindrical lenses and a transmissive $\pi$-step phase plate. The fused silica phase plate has a $882nm$ step oriented at $\pm 40°$ to the grating dispersion direction. A 100μm thick BBO crystal is located 20cm from the output grating. **(b)** TG-SSSI setup, with elements described in the main text. **(c)** Idealized spatiotemporal intensity $I(x,\tau) \propto |E|^2$ and phase $\Delta\Phi(x,\tau)$ of (i) $E_S$ ($l = +1$) described by Eq. (1) and (ii) $E_S^{2\omega}$ ($l = +2$) described by Eq. (2) **(d)** Spatiotemporal intensity and phase of (i) $E_S$ ($l = +1$) and (ii) $E_S^{2\omega}$ ($l = +2$) from simulation [24] of our pulse shaper, followed by SHG in the nonlinear BBO crystal with GVM$_{BBO}$=0 and GDD$_{BBO}$=0. Here, the diamond-like donut shape of $I_S(x,\tau)$ and $I_S^{2\omega}(x,\tau)$ stems from the contribution of higher order Hermite-Gaussian modes generated from the $\pi$-step phase plate and superimposed on the STOV.

Fundamental ($\lambda_0 = 800$ nm) STOVs with electric field $E_S$ were generated by 50 fs pulses from a 1 kHz Ti:Sapphire amplifier routed through the $4f$ pulse shaper depicted in Fig. 1(a), as first presented in refs. [2,7,8]. The key feature of the pulse shaper is the transmissive fused silica phase plate at the common focus of the cylindrical lenses (the Fourier plane of the pulse shaper). The phase plate has a $\pi$ step of height $\lambda_0/2(n_{FS} - 1)$ ~882 nm (where $n_{FS} = 1.4533$ is the refractive index of the substrate at 800nm) across its diameter. Orienting the step at $\alpha = \pm 40°$ to the dispersion direction of the input diffraction gratings generates fundamental STOVs $E_S$ with topological charge $l = \pm 1$ in the near-field of the pulse shaper. Alternatively, a spiral phase plate could have been used to generate a similar STOV in the far field of the shaper [2,8]. The angle $\alpha$ depends on the beam diameter and spectral resolution of the pulse shaper and is tuned experimentally. Second harmonic generation of $E_S$ was accomplished by placing a 100 $\mu$m thick, type I BBO (beta barium borate) crystal at the immediate output of the $4f$ pulse shaper in the near field. The crystal was sufficiently thin to ensure SHG phase matching over the full pulse bandwidth. *It is important to stress here that a well-aligned, ideal pulse shaper of this type* [2,7,8] *imposes the same bandwidth, spatiotemporal phase and polarization on all output photons*. In addition, the photons have a purely *spatial* phase which is related to their extrinsic orbital angular momentum. Minus this component, STOV-based orbital angular momentum is intrinsic, and it is the intrinsic part of the OAM which is responsible for angular momentum conservation under SHG.

In order to observe the spatiotemporal phase and amplitude of the fundamental and SHG STOVs and measure their OAM, we used transient grating single-shot supercontinuum spectral interferometry (TG-SSSI), a technique we developed for pulses containing spatiotemporal phase singularities [2, 22]. TG-SSSI enables single-shot measurement of the phase $\Delta\Phi(x,\tau)$ and intensity $I(x,\tau)$ profiles of ultrashort pulses, where $x$ is a space dimension orthogonal to pulse propagation (as shown in Fig. 1(a)) and $\tau$ local time in the pulse frame. As shown in the TG-SSSI setup depicted in Fig. 1(b), either $E_s$ or its second harmonic $E_s^{2\omega}$ is imaged by a low dispersion $MgF_2$ lens (L1) into the witness plate, where it interferes with a spatial reference pulse $\mathcal{E}_i$ to form a transient volume grating. Spectral interferometry using probe and reference supercontinuum pulses, $E_{pr}$ and $E_{ref}$, is performed on the transient grating, enabling extraction of $\Delta\Phi(x,\tau)$ and $|E_s(x,\tau)|^2 \propto I(x,\tau)$ (or $\Delta\Phi^{2\omega}(x,\tau)$ and $|E_s^{2\omega}(x,\tau)|^2$) [2, 22]. Our TG-SSSI setup can measure pulses as short as ~11 fs at the fundamental (~27 fs at the second harmonic). For adequate signal-to-noise, the lowest pulse energy measured was 3μJ, corresponding to peak intensity ~150 GW/cm$^2$.

A STOV-carrying pulse of centre frequency $\omega_0$ at position $z$ along the propagation axis can be expressed as [1,2]

$$E_s(\boldsymbol{r}_\perp, z, \tau) = a\left(\frac{\tau}{\tau_s} \pm i\,\mathrm{sgn}(l)\frac{x}{x_s}\right)^{|l|} e^{-ik\xi} E_0(\boldsymbol{r}_\perp, z, \tau)$$
$$= A(x,\tau) e^{il\Phi_{s-t}} e^{-ik\xi} E_0(\boldsymbol{r}_\perp, z, \tau), \qquad (1)$$

where $\boldsymbol{r}_\perp = (x,y)$, $\tau = t - z/v_g$ is a time coordinate local to the pulse, $v_g$ is the group velocity, $\xi = v_g\tau$, $\tau_s$ and $x_s$ are temporal and spatial scale widths of the STOV, $\Phi_{s-t}(x,\tau)$ is the space-time phase circulation (spatiotemporal phase) in $x-\tau$ space, $l = \pm 1, \pm 2, \ldots$ is the topological charge of the STOV, $A(x,\tau) = a((\tau/\tau_s)^2 + (x/x_s)^2)^{|l|/2}$, $a = \sqrt{2}((x_0/x_s)^2 + (\tau_0/\tau_s)^2)^{-1/2}$ for $l = \pm 1$, and $E_0$ is the STOV-free Gaussian pulse input to the pulse shaper, where $x_0$ and $\tau_0$ are spatial and temporal widths of the pulse [2]. Here, $a$ is a normalization factor ensuring energy conservation through the pulse shaper. The propagation phase factor $e^{-ik\xi}$ contributes a purely spatial phase shift to the light, contributing to extrinsic orbital angular momentum, and does not contribute to the SHG process.

The process of SHG involving monochromatic and polychromatic beams is well known [23], where, given perfect phase matching and sufficient bandwidth in the undepleted pump regime, the nonlinear polarization and second harmonic field output is proportional to the square of the input field. As applied to the fundamental STOV pulse of Eq. (1), the same process would give

$$E_s^{2\omega}(\boldsymbol{r}_\perp, z, \tau) = A^2(x,\tau) e^{i2l\Phi_{s-t}} e^{-i2k\xi} E_0^2(\boldsymbol{r}_\perp, z, \tau). \qquad (2)$$

Equation (2) predicts that the frequency doubled pulse will have twice the vorticity, topological charge, and angular momentum as the fundamental STOV-carrying pulse. This result is plotted in Fig. 1(c), which shows the intensity and phase of the fundamental (red colormap (i)) and second harmonic fields (blue colormap (ii)). The $2\pi$ phase winding of $E_s$ is transformed into a $4\pi$ phase winding of $E_s^{2\omega}$, accompanied by a narrowing of the intensity ring by a factor $\sqrt{2}$. Because our current pulse shaper modulates only the input pulse phase and not its amplitude, the STOVs it generates are not fully symmetric, as shown in the pulse shaper simulation [24] of Fig. 1(d)(i). The diamond-shaped space-time donut— reproduced in our measurements, as seen later— results from beam contributions by higher order Hermite-Gaussian modes generated at each frequency by the $\pi$-step of the phase plate. The corresponding second harmonic field of the shaper output is shown in Fig. 1(d)(ii).

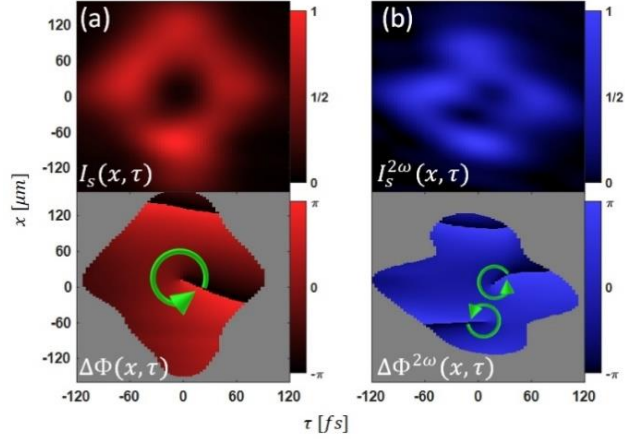

**Fig. 2.** TG-SSSI measurements of fundamental and SHG STOVs. **(a)** Top: Intensity profile $I_S(x,\tau)$ of fundamental $l = +1$ STOV; Bottom: spatiotemporal phase $\Delta\Phi(x,\tau)$ showing one $2\pi$ winding. **(b)** Top: SHG output pulse $I_S^{2\omega}(x,\tau)$ showing two donut holes embedded in pulse; Bottom: spatiotemporal phase profile $\Delta\Phi^{2\omega}(x,\tau)$ showing two $2\pi$ windings. Phase traces are blanked in regions of negligible intensity, where phase extraction fails. These images represent 500 shot averages: the extracted phase shift from each spectral interferogram is extracted, then the fringes of each frame (shot) are aligned and averaged, then the phase map is extracted [22].

The TG-SSSI measurements of the fundamental and SHG STOVs are shown in Fig. 2, where the red colourmap panels of (a) show the spatiotemporal intensity $I_S(x,\tau)$ and phase $\Delta\Phi(x,\tau)$ of the fundamental $l = +1$ STOV $E_S(x,\tau)$ at the near-field output of the $4f$ pulse shaper. $I_S(x,\tau)$ has the characteristic edge-first "flying donut" profile, with the pulse propagating right-to-left, while $\Delta\Phi(x,\tau)$ is a single $2\pi$ winding centred at $(x,\tau) = (0,0)$. The dip in intensity near $x = -60\mu m$ in Fig. 2(a) is due to scattering off the $\pi$-step of the phase plate. Figure 2(b), in blue colormap, shows the measured spatiotemporal intensity $I_S^{2\omega}(x,\tau)$ and phase $\Delta\Phi^{2\omega}(x,\tau)$ of $E_S^{2\omega}(x,\tau)$. Instead of a single $l = +2$ STOV, for which $I_S^{2\omega}(x,\tau)$ would have a single intensity null and $\Delta\Phi^{2\omega}(x,\tau)$ would have a $4\pi$ phase winding (as in Fig. 1(c) and (d)), we see that $I_S^{2\omega}(x,\tau)$ and $\Delta\Phi^{2\omega}(x,\tau)$ show two spatiotemporally offset vortices, embedded in the second harmonic pulse, around whose centers are two $2\pi$ phase windings. This constitutes two $l = +1$ STOVs, and thus energy conservation dictates that the $E_S^{2\omega}$ pulse carries, on average, twice the OAM per photon of the fundamental $E_S$.

The spatiotemporal splitting of the STOV in $E_S^{2\omega}$ is due to (1) group velocity mismatch GVM ($= 1/v_g^{(2\omega)} - 1/v_g^{(\omega)}$) between the $E_S^{\omega}$ and $E_S^{2\omega}$ pulses in the BBO crystal [25] and (2) group delay dispersion (GDD) in both the BBO and lens L1. This is demonstrated by spectral domain ($\tau \to \omega$ and $\mathbf{r}_\perp \to \mathbf{k}_\perp = (k_x, k_y)$) simulations using the carrier resolved unidirectional pulse propagation equation (UPPE) algorithm [26], of the 3D system of propagation equations

$$\frac{\partial \widetilde{E}}{\partial z} = iK_z(\omega, \mathbf{k}_\perp)\widetilde{E} + \frac{i2\pi}{K_z(\omega, \mathbf{k}_\perp)}\frac{\omega^2}{c^2}\widetilde{P}, \qquad (3)$$

for the fields $\widetilde{\mathbf{E}} = \hat{\mathbf{x}} E_S$ or $\widetilde{\mathbf{E}} = \hat{\mathbf{y}} E_S^{2\omega}$. Here, $K_z(\omega, \mathbf{k}_\perp) = \sqrt{k^2(\omega) - |\mathbf{k}_\perp|^2}$ is the linear propagator in the spectral domain, $k(\omega) = \omega n(\omega)/c$ is the wavenumber (with dispersion in BBO and MgF$_2$ lens L1 provided by refs. [25] and [27]) and $\widetilde{P}$ is the nonlinear polarization for the BBO portion of the propagation, where the orthogonally polarized $E_S^{\omega}$ and $E_S^{2\omega}$ fields are computed in the spatiotemporal domain and coupled through

$$P_x = \chi^{(2)}_{xyx}(-\omega; 2\omega, -\omega)E_y E_x^* e^{-i(k_x+k_x-k_y)z}, \quad (4a)$$

$$P_y = \frac{1}{2}\chi^{(2)}_{yxx}(-2\omega; \omega, \omega)E_x^2 e^{i(k_x+k_x-k_y)z}, \quad (4b)$$

where $\chi^{(2)}$ is the second order susceptibility tensor for BBO [28].

Owing to symmetry along y, we used $\partial \tilde{\mathbf{E}}/\partial y = 0$, which also reduces the computational load. The simulation [24] propagates the $E_S$ and $E_S^{2\omega}$ fields through the shaper, through the BBO crystal, to the MgF$_2$ lens L1, and then to the volume occupied by the witness plate. The initial conditions at the entrance to the pulse shaper are $\hat{\mathbf{y}} E_0 = 0$ and $\hat{\mathbf{x}} E_0$ is a plane wave with wavevector $(0,0,k_z)$, where $|E_0|^2$ is taken to be a Gaussian corresponding to the experiment's 3.2 mm $1/e^2$

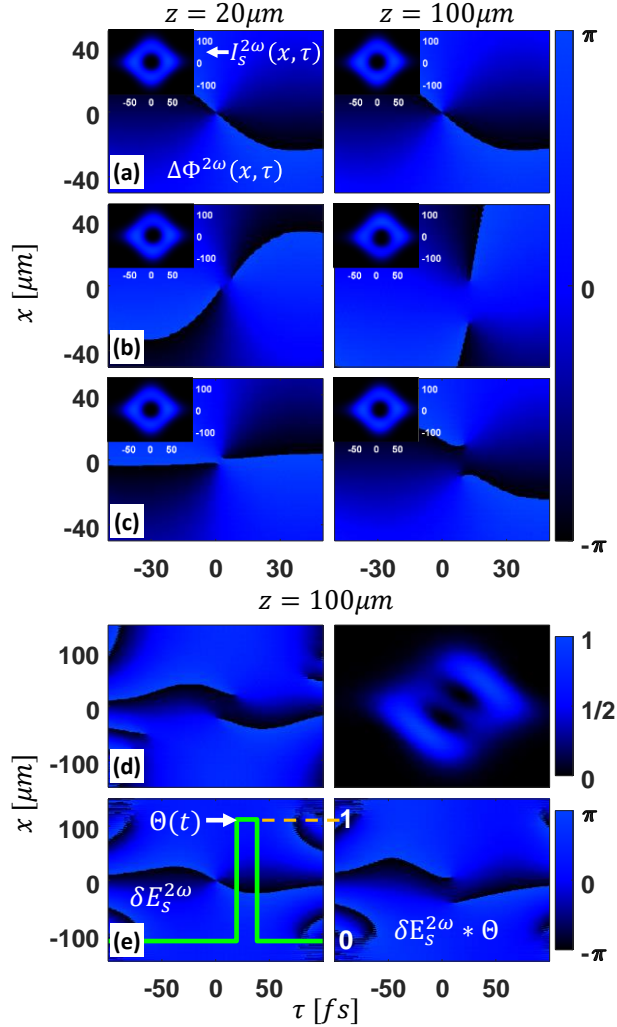

**Fig. 3**. Simulation [24] of the SHG of a fundamental $l = +1$ STOV $E_S$. Plotted is $I_S^{2\omega}(x,\tau) \propto |E_S^{2\omega}|^2$ and $\Delta\Phi^{2\omega}(x,\tau)$, the intensity and phase of the simulated SHG pulse. The simulation generates $E_S$ in the pulse shaper, propagates it through the BBO while generating $E_S^{2\omega}$, and then propagates the fields through MgF$_2$ lens L1 to the witness plate. The phase images are zoomed in at the null to show the decomposition of the high order STOV. **(a)** GVM$_{BBO}$=0, GDD$_{BBO}$=0, GDD$_{L1}$=0. The $l = 2$ STOV does not break up regardless of the BBO thickness, here for 20µm (left) and 100µm (right). Insets: $I_S^{2\omega}(x,\tau)$, with horizontal $\tau(fs)$ and vertical $x$ ($\mu m$) scales. **(b)** GVM$_{BBO}$ = 0.19 $fs/\mu m$, GDD$_{BBO}$=0, and GDD$_{L1}$=0. Here, the $l = 2$ STOV breaks up into two $l = 1$ STOVs in as little as 20µm of propagation in BBO. **(c)** GVM$_{BBO}$ = 0.19 fs/µm, GDD$_{BBO}$=20.9 fs, and GDD$_{L1}$=0. **(d)** Simulation corresponding to experimental parameters of Fig. 2: GVM$_{BBO}$=0.19 fs/µm, GDD$_{BBO}$=20.9 fs, and GDD$_{L1}$ = 350 fs$^2$. **(e)** Simple convolution model of the effect of GVM in BBO, showing decomposition of $l = +2$ STOV into two $l = +1$ STOVs.

beam radius, 50 fs pulsewidth and 350 µJ energy.

Our simulations generating $E_S^{2\omega}(\mathbf{r}_\perp, z, \tau)$ show that for the case of zero dispersion ($\text{GVM}_{\text{BBO}} = 0$, $\text{GDD}_{\text{BBO}} = 0$, and $\text{GDD}_{\text{L1}} = 0$), the $l = +2$ STOV does not break up for BBO crystal thickness less than $\sim 100 \mu m$. This is seen in Fig. 3(a) for two BBO thicknesses, 20 µm and 100 µm. The phase plots for $\Delta\Phi^{2\omega}(x,\tau)$ are zoomed in near the phase singularity, while the insets show the full intensity profile.

Figure 3(b) shows a simulation for the case of $\text{GVM}_{\text{BBO}} = 0.19$ fs/µm and $\text{GDD}_{\text{BBO}}$ =0. We conclude that non-zero $\text{GVM}_{\text{BBO}}$ is sufficient to break the $l = +2$ STOV into two $l = +1$ STOVs in as little as 20 µm of propagation. Figure 3(c) shows the result of including $\text{GDD}_{\text{BBO}} = 20.9$ fs in the simulation, which slows the separation of the two windings relative to the case in Figure 3(b). We note that in the spatiotemporal intensity profiles $I_S^{2\omega}(x,\tau)$ (in the insets of (a), (b), and (c)) the two zeros of the field are resolved only into one central intensity null due to their proximity to each other.

A simulation corresponding directly to Fig. 2's experimental parameters is shown in Fig. 3(d), where $\text{GDD}_{\text{L1}} = 350$ fs$^2$ is also added to the nonzero $\text{GVM}_{\text{BBO}}$ and $\text{GDD}_{\text{BBO}}$. Here, the already separated $l = +1$ STOVs are driven farther apart by the additional $\text{GDD}_{\text{L1}}$, leading to two spatiotemporally offset nulls in $I_S^{2\omega}(x,\tau)$. Because of linear propagation in L1, $\text{GDD}_{\text{L1}}$ has a different effect than the interplay of $\text{GVM}_{\text{BBO}}$ and $\text{GDD}_{\text{BBO}}$ on vortex separation during nonlinear propagation (Fig. 3(c)). Comparing the results in Figs. 2(b) and 3(c), the simulation matches the experiment quite well.

Physically, the effect of group velocity mismatch in the BBO on the decomposition of an $l = +2$ STOV can be explained as follows: As $E_S$ propagates in the BBO, each portion of its envelope at local time $\tau$ nonlinearly generates a contribution $\delta E_S^{2\omega}$ which slips back in time. The SHG crystal output can then be constructed as the convolution $E_S^{2\omega}(x,\tau) = \int_{-\infty}^{\tau} \delta E_S^{2\omega}(x, \tau - t)\Theta(t)\, dt$, the sum of a sequence of time-shifted $l = +2$ STOV contributions $\delta E_S^{2\omega}(x, \tau - t)$ that models the increasing slip of the peak of $E_S^{2\omega}$ with respect to the peak of $E_S$. Here $\Theta(t) = 1$ for $0 \le t \le \Delta\tau$ and $\Theta(t) = 0$ elsewhere, where the maximum time slip is $\Delta\tau = (1/v_g^{(2\omega)} - 1/v_g^{(\omega)}) L \approx 19$ fs for the SHG crystal length $L = 100\mu m$. The integral yields two spatially offset $l = +1$ STOVs, as depicted in Fig 3(d). This is essentially the STOV equivalent to the splitting observed due to spatial walk-off of LG beams in nonlinear crystals [29]. We note that the addition of nonzero $\text{GDD}_{\text{L1}}$ leads to the diagonal (spatiotemporal) offset of Fig. 3(c). Recognizing from Fig. 3 that the two spatiotemporally offset $l = +1$ STOVs represents a superposition of time-shifted $l = +2$ STOV pulses, we find that OAM conservation in second harmonic generation also applies to STOVs.

In summary, we have experimentally and theoretically demonstrated the conservation of STOV-based OAM in second harmonic generation. Group velocity mismatch between the fundamental and second harmonic STOVs is the primary cause for $l = +2$ STOVs to quickly separate into two $l = +1$ STOVs after only a short propagation distance in the SHG crystal. In general, once a higher order STOV with $|l| > 1$ is generated, spatiotemporal decomposition of vorticity is driven by diffraction and dispersion in propagation media. In particular, the space-time separation of STOVs during SHG could be mitigated via group velocity matching by using non-collinear SHG geometry.

The question of whether photons in an ultrashort STOV pulse individually carry transverse OAM is difficult to answer experimentally in linear optics: one would need to attenuate the pulse and somehow examine the statistical spatiotemporal distribution of photons. However, this question is more easily answered using nonlinear optics. The conservation of photon number implied by the Manley-Rowe relations for SHG, $2\, d/dz\, (I^{(\omega)}/\hbar\omega) = d/dz\, (I^{(2\omega)}/\hbar\omega)$

[23], implies that, *on average*, photons at the second harmonic carry twice the OAM of photons at the fundamental. However, because SHG is fundamentally a quantum mechanical process involving light-matter interactions of the quantized electromagnetic field, and because all photons in the STOV pulse from our pulse shaper carry the same bandwidth, polarization, and spatiotemporal phase, we conclude that energy and angular momentum conservation in the SHG process holds at the individual photon level. Therefore, we conclude that photons in STOV-carrying pulses have OAM orthogonal to their direction of propagation.

It is worth noting that if we take a STOV-carrying pulse of the type generated here, and greatly attenuate it to just a few photons, the uncertainty relations $\Delta k_x \Delta x \geq 1/2$ and $\Delta k_\xi \Delta \xi \geq 1/2$ ensure that a photon with STOV OAM could be found anywhere in the transverse and longitudinal extent of the pulse (except at the field null), and could have any frequency consistent with the bandwidth.

The authors thank I. Larkin and R. Schwartz for discussions and technical assistance. This research is supported by the Air Force Office of Scientific Research (FA9550-16-1-0121, FA9550-16-1-0284), the Office of Naval Research (N00014-17-1-2705, N00014-20-1-2233), and the National Science Foundation (PHY2010511).

# Second harmonic generation of spatiotemporal optical vortices (STOVs) and conservation of orbital angular momentum: supplemental material

S.W. Hancock, S. Zahedpour, and H. M. Milchberg

*Institute for Research in Electronics and Applied Physics, University of Maryland College Park, Maryland, 20742, USA*

Propagation simulations of this paper employed the Unidirectional Pulse Propagation Approximation (UPPE) algorithm [1], a fully spectral method. UPPE simulated (1) nonlinear propagation of the fundamental $l = 1$ STOV $E_S$ through the BBO and generation of the SHG field $E_S^{2\omega}$ (2), linear propagation from the exit face of the BBO through lens L1 to the witness plate of TG-SSSI (See Fig. 1 of main paper). UPPE can be implemented in multiple ways, none of which require the beams to be paraxial. The only assumption made is unidirectional propagation. For the simulations of this paper, we implemented a carrier-resolving scalar UPPE solver.

Assuming $\mathbf{E}(x, y, t, z)$ is the electric field of the laser pulse (including the carrier) and $\tilde{\mathbf{E}}(k_x, k_y, \omega, z)$ is its Fourier transform in time ($\tau \to \omega$) and transverse space ($\{x, y\} \to \{k_x, k_y\} = \{\mathbf{k}_\perp\}$) we have

$$\partial \tilde{\mathbf{E}}/\partial z = iK_z(\omega, \mathbf{k}_\perp)\tilde{\mathbf{E}} + i2\pi K_z^{-1}(\omega, \mathbf{k}_\perp)(\omega^2/c^2)\tilde{\mathbf{P}} \qquad (S1)$$

where $K_z(\omega, \mathbf{k}_\perp) = \sqrt{k^2(\omega) - |\mathbf{k}_\perp|^2}$ models diffraction and dispersion and $\tilde{\mathbf{P}}$ is the nonlinear polarization. The advantage of solving spectral domain Eq. (S1) is the decoupling of field values from their adjacent values (in $\omega$ and $\mathbf{k}_\perp$). Only the history of the fields $\tilde{\mathbf{E}}$ and $\tilde{\mathbf{P}}$ at a given spectral point $(\omega, \mathbf{k}_\perp, z)$ determines the value of $(\omega, \mathbf{k}_\perp, z + \Delta z)$ through simple integration. Depending on the type of nonlinearity modeled, it is convenient to calculate the polarization in either the spatio-temporal domain ($\mathbf{P}$) or the spatio-spectral domain ($\tilde{\mathbf{P}}$). For nonlinear propagation through the BBO crystal, $\mathbf{P}$ is calculated in the space-time domain since the pulse is polychromatic, and is given by

$$P_x = \chi^{(2)}_{xyx}(-\omega;\ 2\omega, -\omega) E_y E_x^* \, e^{-i(k_x + k_x - k_y)z}$$
$$P_y = (1/2)\chi^{(2)}_{yxx}(-2\omega;\ \omega, \omega) E_x^2 e^{i(k_x + k_x - k_y)z},$$

while it is $P = 0$ in the linear part of the propagation. For inclusion in Eq. (1) of the main paper, $\mathbf{P}$ is Fourier transformed back to $\tilde{\mathbf{P}}$. To apply our simulations to y-symmetric STOVs we use $\partial \tilde{E}/\partial y = 0$ ($k_y = 0$), thereby reducing the simulations to a 2D slice through the pulse at $y = 0$.

Simulation of the pulse shaper was performed by applying a $\pi$-step phase to a simulated Gaussian pulse in the spatio-spectral domain. The pulse parameters used were: $\tau_{FWHM} = 50\,\text{fs}$, $x_{FWHM} = 4\,\text{mm}$, central wavelength $\lambda_0 = 800\,\text{nm}$, and grating period $\Lambda = 1/1200$ mm. The input field into the pulse shaper is $E_0(x, \tau) = E_0 \exp(-x^2/2\sigma_x^2 - \tau^2/2\sigma_\tau^2 + i\omega_0\tau)$, where $\sigma_x = x_{FWHM}/2\sqrt{\ln(2)}$, $\sigma_\tau = \tau_{FWHM}/2\sqrt{\ln(2)}$ and $\omega_0 = 2\pi c/\lambda_0$. An input grating angle of $\theta_i = -10°$ was assumed and the first order diffraction angle is given as $\theta_1(\lambda) = \sin^{-1}(-\lambda/\Lambda - \sin\theta_i)$.

Referenced to $\lambda_0$ in the first order, the angle is $\Delta\theta_1(\lambda) = \theta_1(\lambda) - \theta_1(\lambda_0)$. The spatial location of the dispersed ray at the π-step phase plate is $y = f \tan^{-1} \Delta\theta_1(\lambda)$, where $\lambda = 2\pi c/\omega$, and $f = 100mm$ is the focal length of the cylindrical lens. The phase plate is rotated at an angle $\alpha$ relative to the dispersion direction of the grating, and the phase shift it imparts to the beam is computed as $A(x,y) = 0$ for $x < y \tan\alpha$ and $A(x,y) = \pi$ for $x > y \tan\alpha$, where $E_{out}(x,\tau) = \mathcal{F}_\omega^{-1}[\mathcal{F}_\tau[E(x,\tau)] \exp(iA(x,y))]$. Simulating the scattering off of the π-step on the phase plate was similarly done by defining a mask $B(x,y) = 0$ for $|x - y \tan\alpha| \leq w/2$ and $B = 1$ elsewhere, where $w$ is the width of the scattering edge (the etched step with finite width in the phase plate). The scattering (amplitude) mask is then applied to the input pulse as $E_{out}(x,\tau) = \mathcal{F}_\omega^{-1}[B(x,y)\mathcal{F}_\tau[E(x,\tau)] \exp(iA(x,y))]$. Figure S1 shows the simulated shaper output without and with the scattering edge included in the simulation for a range of $\alpha$ values. The pulse where the scattering edge is simulated show a slight dip in intensity at the bottom of the pulse due to amplitude modulation in the spatio-spectral domain.

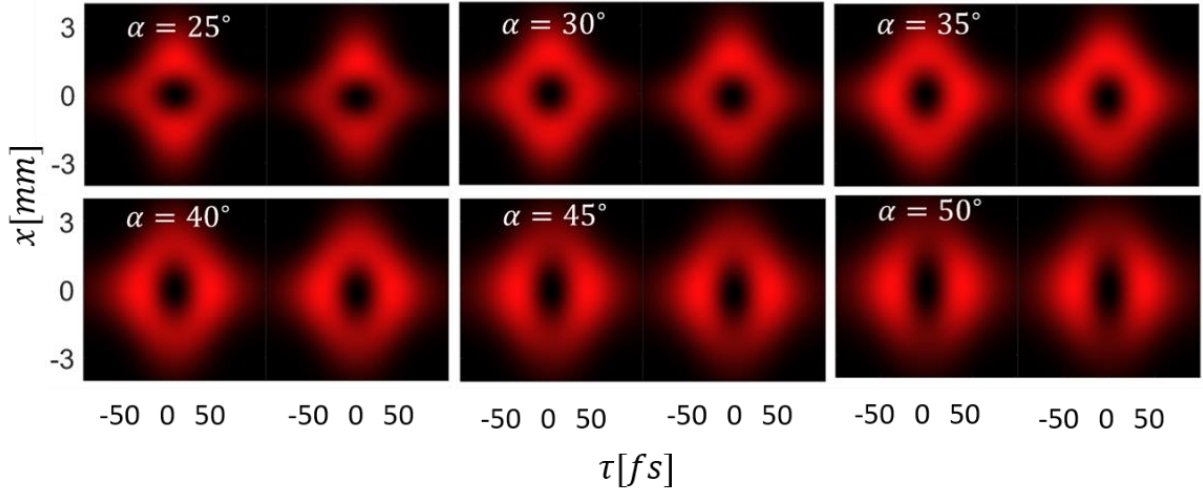

**Figure S1.** Intensity of the STOV at the pulse shaper output vs. π-step angle ($\alpha$) with respect to dispersion direction (y). For each value of $\alpha$, the left panel does not include scattering due to the π-step, while the right panel includes it by applying the phase mask B discussed above.

The diamond-like shape of the simulated shaper output of Fig. S1 is a result of high order Hermite-Gaussian modes generated at each frequency bisected by the π-step of the phase plate. To illustrate this Fig. S2(a) shows the intensity output of the pulse shaper at $\alpha = 40°$ (without the scattering edge), while Fig. S2(c) shows $\log(|\mathcal{F}_x(E_{out})| + 1)$. The high frequency components in the spectral domain of the field can be seen as horizontal strips. Figure S2(b), shows the result of spatially filtering the output of the pulse shaper (the spatial filtering here was performed by setting $k_x > 2.36mm^{-1} = 0$ in (c)). Finally, Fig. S2(d) shows $\log(|\mathcal{F}_x(E_{out})| + 1)$ for an ideal symmetric STOV of the form $(\tau/\tau_s \pm i x/x_s)E_0(r_\perp, z, \tau)$. We note the absence of the horizontal bands in the spectrum.

Propagation in free-space of STOV was previously explored and published [2]. Figure S3 shows the intensity and phase evolution during propagation of the simulated pulse shaper output and an ideal STOV over the Rayleigh range of the Gaussian input pulse using the previously described UPPE method. Note that both ideal and shaper generated STOV evolve into two-lobed structures after one Rayleigh length of propagation.

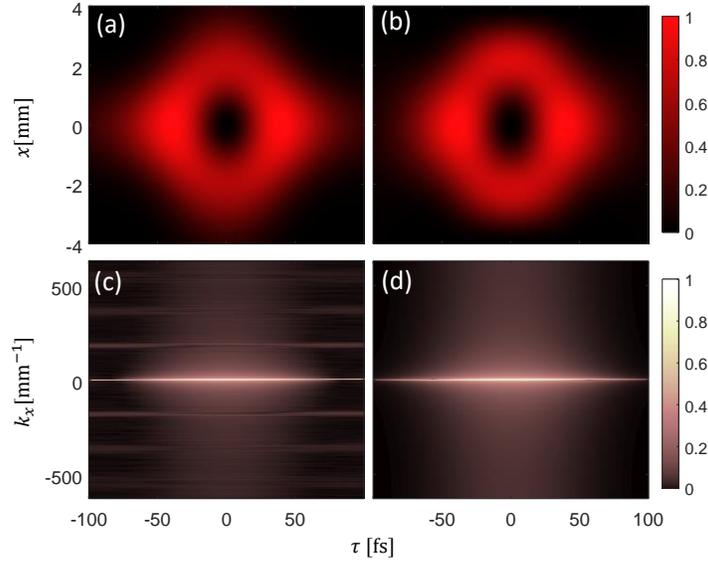

**Figure S2.** (a) Pulse shaper output for $\alpha = 40°$ without modeling scattering from edge of the $\pi$-step phase plate. (b) shows the spatially filtered output of the pulse shaper where large values of $k_x$ in (c) are set to zero. (c) shows the log of the spatial Fourier transform of the shaper output, $\log(|\mathcal{F}_x(E_{out})| + 1)$. The high frequency horizontal stripes can be filtered out using a spatial filter which result in (b) the spatially filtered output of the pulse shaper (d) shows the log of the spatial Fourier transform of an ideal STOV, $\log(|\mathcal{F}_x(E_{STOV})| + 1)$.

Additionally, we show the intensity and phase evolution of over the free-space propagation of the

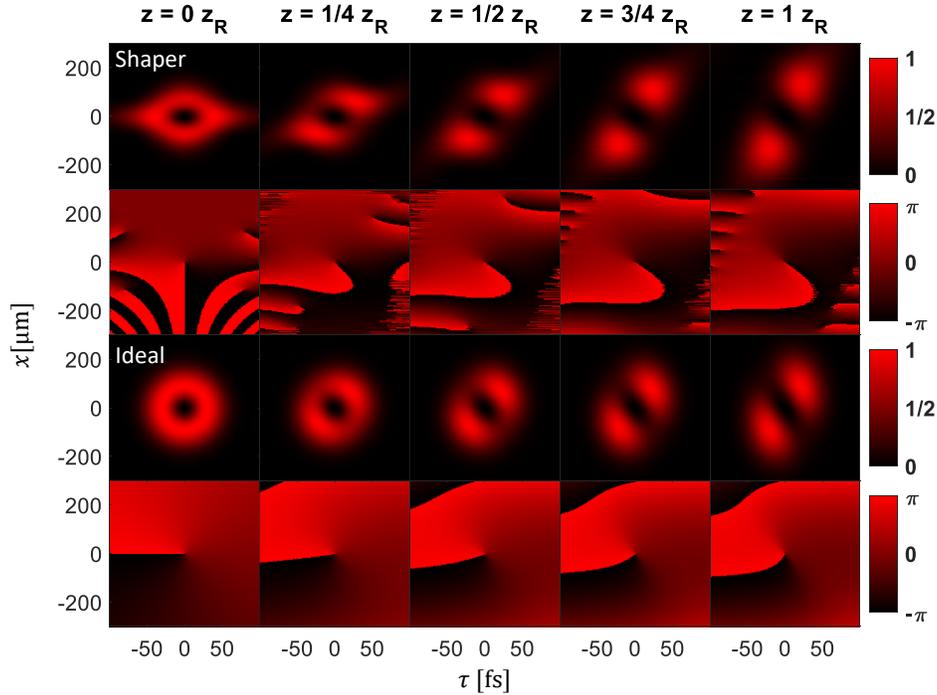

**Figure S3.** (Shaper) shows the intensity (top) and phase (bottom) of the shaper generated STOV pulse as it propagates over one Rayleigh length, $z_R$, of the Gaussian carrier pulse. (Ideal) shows the intensity (top) and phase (bottom) of an ideal STOV pulse as it propagates over one Rayleigh length, $z_R$, of the Gaussian carrier pulse.

squared output of the pulse shaper (second harmonic generation in an ideal crystal), and for that of an ideal STOV over the Rayleigh range of the frequency double Gaussian input pulse in Figure

4. Note that the second harmonic of both the ideal STOV and the shaper-generated STOV evolve into three-lobed structures after propagation of one Rayleigh length.

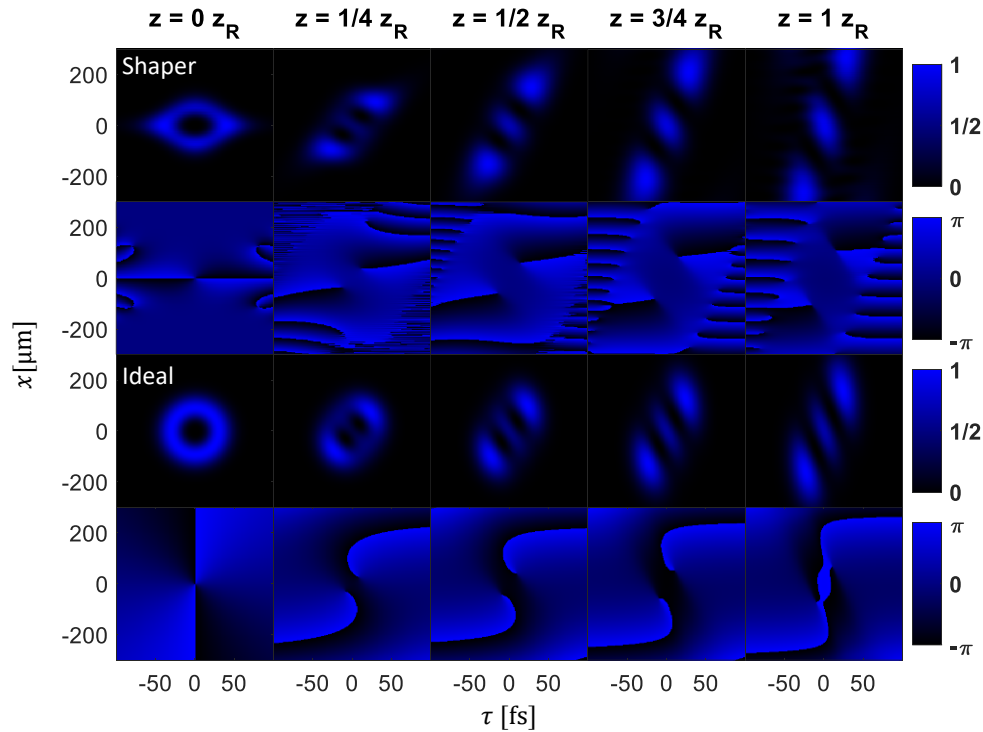

**Figure S4.** (Shaper) shows the intensity (top) and phase (bottom) of the second harmonic of the shaper generated STOV pulse as it propagates over one Rayleigh length, $z_R$, of the second harmonic of Gaussian carrier pulse. (Ideal) shows the intensity (top) and phase (bottom) of the second harmonic an ideal STOV pulse as it propagates over one Rayleigh length, $z_R$, of the second harmonic of Gaussian carrier pulse.